# RECENT ADVANCES IN SOLID-STATE ORGANIC LASERS

**Sébastien Chénais and Sébastien Forget**

Laboratoire de Physique des Lasers, Institut Galilée

Université Paris 13 / C.N.R.S.

99 avenue J.-B. Clément, 93430 Villetaneuse, France

*Organic solid-state lasers are reviewed, with a special emphasis on works published during the last decade. Referring originally to dyes in solid-state polymeric matrices, organic lasers also include the rich family of organic semiconductors, paced by the rapid development of organic light emitting diodes. Organic lasers are broadly tunable coherent sources are potentially compact, convenient and manufactured at low-costs. In this review, we describe the basic photophysics of the materials used as gain media in organic lasers with a specific look at the distinctive feature of dyes and semiconductors. We also outline the laser architectures used in state-of-the-art organic lasers and the performances of these devices with regard to output power, lifetime, and beam quality. A survey of the recent trends in the field is given, highlighting the latest developments in terms of wavelength coverage, wavelength agility, efficiency and compactness, or towards integrated low-cost sources, with a special focus on the great challenges remaining for achieving direct electrical pumping. Finally, we discuss the very recent demonstration of new kinds of organic lasers based on polaritons or surface plasmons, which open new and very promising routes in the field of organic nanophotonics.*

## 1 INTRODUCTION

Since the first demonstration of laser oscillation, now more than fifty years ago[1], applications using lasers have spread over virtually all areas, *e.g.* research, medicine, technology or telecommunications. The variety of available laser sources is large and covers a large span in terms of attainable wavelengths, output powers and pulse duration. Although they can have different aspects, lasers are always composed of three essential building blocks (see fig. 1): a gain medium, a pump source and a resonator. The gain medium can be a gas, a liquid or a solid in which amplification of optical waves occurs by stimulated emission of radiation. This fundamental process is dominant over absorption provided that a population inversion is realized, which is the role attributed to the pump source. The resonator provides feedback and defines the spatial and spectral coherence of the beam[2].

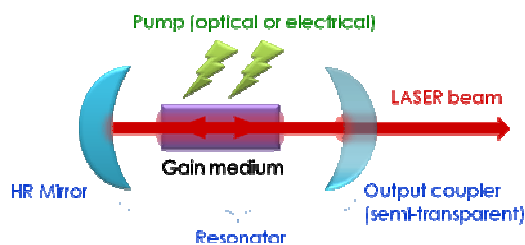

**Figure 1** : *the essential building blocks of a laser : a gain medium, a resonator (here, in its genuine form, two curved mirrors including a semi-transparent outcoupler) and a pumping source.*

Among all kind of laser materials, π-conjugated molecules or polymers are of foremost interest and form the rich family of so-called "organic lasers". The history of organic lasers is almost as long as the story of lasers themselves : liquid dye lasers, based on solutions of π-conjugated highly luminescent (dye) molecules, were demonstrated as early as in the mid-sixties[3], and are still today popular sources of tunable visible radiation, in spite of their cumbersome designs and the inconvenience linked to toxic solvents. Very early (as soon as 1967) it was proposed to incorporate dyes in solid-state polymeric matrices[4] which appeared to be a promising route to build broadly tunable sources that would have the benefit to be compact, convenient and manufactured at low-costs. Although there is still today active research on dye-doped polymer lasers, these devices did not manage, from then and up to now, to enter the marketplace: the main reason is probably that there is a fundamental contradiction between the inherent bad photostability of organic molecules and the high cost of the pump source, in general a pulsed laser. Subsequent efforts then naturally went in the direction of improving the dye and host matrix photostability, and/or decreasing the required pump threshold intensity so that alternatives to pulsed lasers may be found. The history of solid-state organic lasers reached its turning point with the advent of organic semiconductors, paced by the development of organic light emitting diodes[5]. Very-low thresholds could be demonstrated in thin-film based organic semiconductor lasers; furthermore the devices turned out to be easy-to-handle and compact, while keeping all the advantages of organic materials [6, 7]. The question was then whether an *organic laser diode* would be ever realized, that is a device pumped with an injected current rather than with an optical source, in the same way as inorganic semiconductor lasers work. This issue drove considerable efforts and is still now a major inspiration railroad for organic laser research, although no demonstration of such a device has been reported yet.

What are the targeted applications for organic solid-state lasers? The natural playground associated to organic lasers is spectroscopy. In this case, the modest output power is not problematic, whereas the broad tunability is a strong advantage and short pulses allow time-resolved studies. The simplicity of the fabrication process makes organic sources attractive for integration onto miniature spectroscopic systems. In a similar way, organic lasers are potentially useful for chemical sensing. An interesting achievement was the detection of trinitrotoluene (TNT) using conjugated polymer films[8]. The mechanism is based on the observation of fluorescence quenching which results from the presence of a TNT molecule on the polymer. Even if this detection is possible with simple fluorescence, it has been demonstrated that the sensitivity is much higher when working with lasers operating around threshold[9]. Conjugated polymers could be used in a similar way to detect specific DNA sequences or metal ions[10, 11]. The easy integration of organic solid-state lasers makes then ideal sources for lab-on-a-chip sensors for biophotonics, coupled to microfluidic devices [12]. Organic semiconductor lasers may also find their place in data communications. Obviously, nobody can seriously think of competing with silica optical fibers to carry optical signals for long-range communications; but conversely short-haul data transfer (for example in cars or Fiber-To-The-Home transfer) can take benefit of polymer fibers, essentially because of their much lower cost. Optical amplifiers based on organic media are thus a simple and compatible technology that can be used together with polymer fibres within their transmission windows (around 530, 570 and 650 nm)[13].

In this paper, organic solid-state lasers are reviewed. The reader will find several very comprehensive and top-quality reviews on that subject published all along the last decade [14-22]. Therefore, this paper is not intended to be a complete survey of every aspect of organic lasing; it is instead written with the purpose of drawing a picture of the field as it appears to us at the beginning of the year 2011, emphasizing on the latest advances and trends, particularly in the field of device physics.

This paper is structured as follows: in **Section 2**, we briefly review the different types of organic solid-state materials used for lasers, and comment on the distinction that can be made between a "dye "and an "organic semiconductor" from a laser physicist point of view. **Section 3** will be devoted to an overview of existing organic laser architectures, with a special emphasis on devices employing thin films. In the next sections (4 to 6), we attempt to analyze the recent research trends. Relying on papers published during the last decade, a few general tendencies can be sketched. First of all, the quest of the "organic laser diode" remains the top keyword that drives the scientific community. Many insights contributed to a better understanding of the bottlenecks caused by the supplementary losses brought by electrical excitation; surprisingly though, some claims of apparent gain or lasing action[23] under electrical excitation have been published, but they were later attributed to artefacts[24] or mistaken identification of laser action[25]. These aspects will be reviewed in **Section 4**. In the meantime, new and alternative concepts came to birth. Optically-pumped organic lasers ceased to be regarded as intermediate steps before achieving a future — maybe unlikely? — electrically-pumped device; instead they started to be regarded as interesting devices *on their own*, as far as cheap pump sources, such as diode-pumped solid-state microlasers[26], laser diodes[27] or even light-emitting diodes[28], could be used. In **Section 5** of this report, we will highlight those recent progresses made in this direction with optically-pumped organic lasers, either in terms of lowering the threshold, increasing the wavelength coverage (to the deep red[29] or IR and down to the UV[30]), improving the wavelength agility, enhancing the lifetime of the devices, or improving the conversion efficiency, output power and beam quality. The last part (**section 6**) will be devoted to the newest, and maybe the most unexpected trend in the field: using organic gain media in devices that are not classical lasers anymore, or "photon lasers", but rather devices based on *quasiparticles* such as exciton-polaritons (mixing of photon states and excitons) or Surface Plasmons (and Surface Plasmon Polaritons). Two recent breakthroughs illustrate this tendency: the demonstration of an organic-crystal microcavity polariton laser[31], which may open a novel path towards electrically-driven lasers, and the first experimental proof of an organic-based "spaser"[32] (an acronym for "Surface Plasmon Amplification by Stimulated Emission of Radiation"). The latter represents a real step forward as it breaks the supposedly "unbreakable law" stating that the minimum size for a laser has to be half an optical wavelength. Being a truly nanometric — a few tens of nm in size — coherent quantum generator at optical frequencies, the spaser may play the same role in nanophotonics than the laser is playing in photonics, and reconcile the length scales of electronics and optics. These recent results open very exciting perspectives in the use of organic gain media both for fundamental physics and practical applications.

## 2 MATERIALS

### 1. OVERVIEW OF MOLECULAR MATERIALS SUITABLE FOR ORGANIC LASING

Dye molecules dissolved in liquid solvents have been for decades the winning scheme for organic lasers[3]. The first attempts to make solid-state organic lasers naturally consisted in using the same compounds. A "dye" can be defined as a π-conjugated molecule with a high quantum yield of fluorescence, it can be either neutral or ionic: typical examples are xanthenes (rhodamine and fluorescein families), coumarines, oxazines, or pyrromethenes[33]. Even though a given dye molecule can be a very good light emitter in a dilute form, it will be less efficient or even non-emissive at all at high concentrations[33] because of intermolecular interactions arising between nearby molecules, an effect referred to as "concentration quenching". In the solid state, a straightforward strategy to avoid quenching is to incorporate dye molecules as dopants into solid matrices. Solid hosts can be either polymers — essentially poly(methyl methacrylate) (PMMA) and its derivatives[34] — or glasses and hybrid organic-inorganic materials prepared by sol-gel techniques[35-38].

Later, organic semiconductors (OSC) have been identified as suitable for lasing. The term refers to their electrical semiconducting properties, although the physics of electrical conduction in organics differs notably from that of inorganic semiconductors[39]. Based on structural considerations, one can distinguish three categories of organic semiconductors: organic crystals, small molecules and polymers.

Organic single crystals of anthracene or pentacene[40, 41] resemble inorganic crystals, and their transport properties can be defined in terms of bands. In practice, the high voltage needed to get light from those crystals limited their practical interest and their difficult growth limited their lasing applications as well. In addition, high molecular-crystal packing provides high mobilities which is generally in contradiction with a high quantum yield of fluorescence[42]. Single crystals are however being considered for light-emitting transistors[43], which may be of interest for organic lasers as discussed in section 4.

Small-molecular organic semiconductors are at the origin of the astonishing development of commercial thermally-evaporated OLEDs[5, 44]. In such OSCs, which are mostly amorphous, transport occurs by hopping between localized sites, with a probability that critically depends on the orbital intermolecular overlap. Any neat film of a π-conjugated compound will then have semiconducting properties to some extent. However the tendency to intermolecular π-π stacking and dipole-dipole interactions between units leads *a priori* to severe quenching and is detrimental for lasing: many organic semiconductors turn out to be either not emissive or may emit fluorescence but are not suitable for lasing. Many efforts in molecular design have hence been directed towards improving the luminescence properties of organic semiconductor films. This can be done by designing molecular geometries where fluorophore units are kept apart and/or π-π stacking highly reduced[45-47]. Dendrimers for instance are good candidates for organic lasing: the dendrimer structure[48, 49] consists of a chromophore (for example pyrene[50]) located in the core of a highly branched structure ended with surface groups. The core defines the optical properties such as emission wavelength, whereas the surface groups confer solubility and act as spacers to limit π-π stacking. Similarly, the spiro-linkage used to couple two oligomers in spiro-compounds[51-53] enables defining a twisted geometry[54] in which the optical properties of the individual moieties are conserved. However the easiest way to get rid of concentration quenching is still to physically separate the emissive units; it is possible however to give to the host an active role with guest/host systems. The archetypal example is $Alq_3$:DCM[55, 56]. In this case, the pump light is absorbed by the higher energy gap host material ($Alq_3$ – see fig.2.a), which very efficiently transfers its energy to the lower energy gap guest (DCM – see fig.2.d) through a Förster mechanism[57] (figure 3). Compared to DCM doped into a passive matrix (like PMMA), the host/guest structure enables a much higher absorption of the pump; the obtained material is furthermore able to transport charges through $Alq_3$. At last, the Stokes shift between absorption and emission wavelength is increased, which consequently reduces self-absorption (*i.e.* re-absorption of the emitted laser light).

The other major category of semiconducting organic molecules is formed by conjugated polymers[58]. Conjugated polymers can exhibit high photoluminescence yields even in the solid-state. Lasing has been demonstrated in a large number of such macromolecules, especially in poly(phenylene vinylene) (PPV)[18, 59, 60] and polyfluorenes[61, 62] (figure 2.b and 2.c). While it is generally difficult to make homogeneous blends of polymers, host/guest systems composed of two polymers have been recently investigated for laser action, based on MEH-PPV doped into F8BT[63] (figure 2.e and 2.f).

Recently, new materials emerged as good candidates for organic lasing: this is the case of liquid crystals [64], whose self-organized structure forms photonic bandgaps that can be exploited for lasing; the gain unit can be a classical dye incorporated in the structure or it might be the liquid crystal itself [65]. A more exotic lasing material was newly reported by M. Gather *et al.*, who realized a single-cell "living" laser [66], in which the gain medium is Green Fluorescent Protein, a well-known imaging tracer in biophotonics. In addition of its future possible implications in bio-imaging, it presents some new and intriguing aspects, such as the fact that the cell remains alive during laser operation, and the (hypothetical) possibility to create a biological self-generating and self-healing laser medium[67].

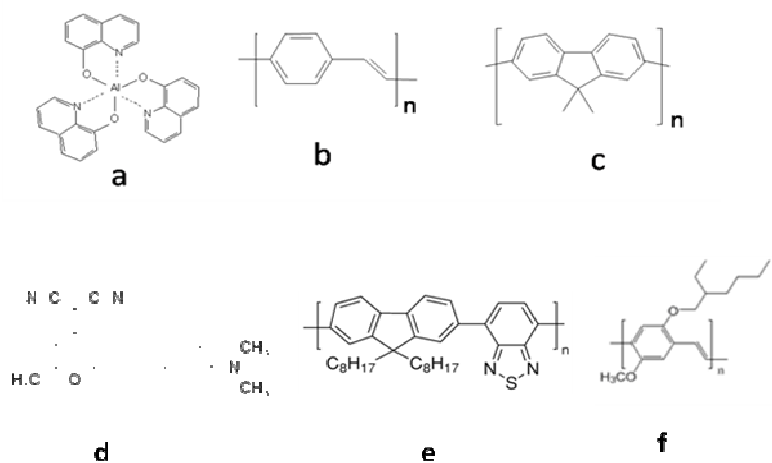

**Figure 2** : Chemical structure of some typical organic media used in organic lasers : (a) aluminium tris(quinolate) or $Alq_3$ (b) generic polyphenylene vinylene or PPV (c) generic polyfluorene (d) DCM (4-(Dicyanomethylene)-2-methyl-6-(4-dimethylaminostyryl)-4H-pyran (e) F8BT [poly(9,9 '-dioctylfluorene-co-benzothiadiazole)] (f) MEH-PPV [poly(2-méthoxy-5-(2-éthyl-hexyloxy)-1,4-phenylene-vinylene)]

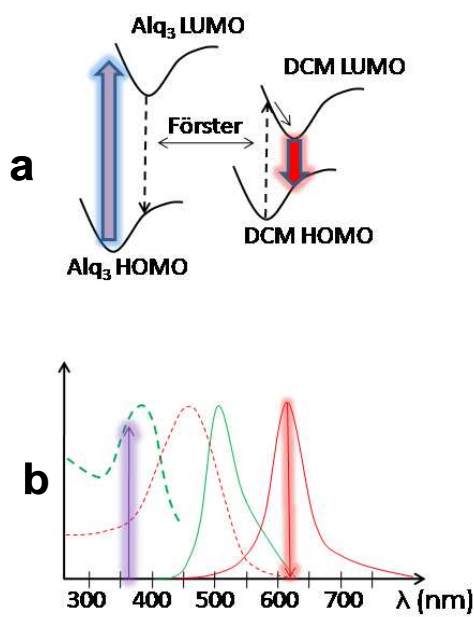

**Figure 3** (a) Illustration of the host-guest Förster energy transfer between $Alq_3$ and DCM and (b) corresponding absorption (dashed line) and emission (full line) spectra. The $Alq_3$ host absorbs a UV pump photon and transfers its excitation through nonradiative dipole-dipole interaction to the guest ; Förster energy transfer is efficient if the overlap between the emission spectrum of the host and the absorption spectrum of the guest is large. It results in a photon emission by the guest at a wavelength where the host absorption is almost zero, thus minimizing reabsorption losses.

## 2. PHOTOPHYSICAL PROPERTIES RELEVANT FOR LASER ACTION

Many photophysical aspects of π-conjugated systems are relevant for lasing. Conjugated organic molecules are intrinsically four-level systems (figure 4): the excited state (or manifold) $S_1$ and the ground state $S_0$ are both coupled to a multitude of vibronic states: the absorption line is between the ground state of the $S_0$ manifold and a vibrational high-lying state of the $S_1$ manifold, while the laser transition occurs between the lowest-energy vibrational state of the $S_1$ manifold and one of the vibrational states of $S_0$. This four-level system guarantees low-threshold lasing since no minimum pump power is required (in addition to cavity losses) to achieve population inversion, provided that the lower level of the laser transition is above the ground state by several times kT [68]. An important aspect related to lasing is the presence of triplet states (see fig. 4). Molecules in the first singlet excited state $S_1$ may flip their spin through intersystem crossing and end up in the $T_1$ triplet state, generally lower in energy, where they can accumulate because of a high triplet-state lifetime.

Unfortunately, the $T_1 \rightarrow T_2$ absorption band is very broad and overlaps the singlet fluorescence emission band[69]. Hence, solid-state organic lasers can indeed only emit *short pulses*, (contrary to liquid dye lasers which can run in CW mode because the continuous circulation of the fluid allows refreshing the medium permanently.) Recent reports have shown the termination of lasing after a few ns[63] in polymer guest/host systems or a few tens of ns[70] in small-molecular lasers after pump turn-on in case of long-pulse pumping: these aspects are essential in the context of electrical pumping and are reviewed in more detail in section 4.

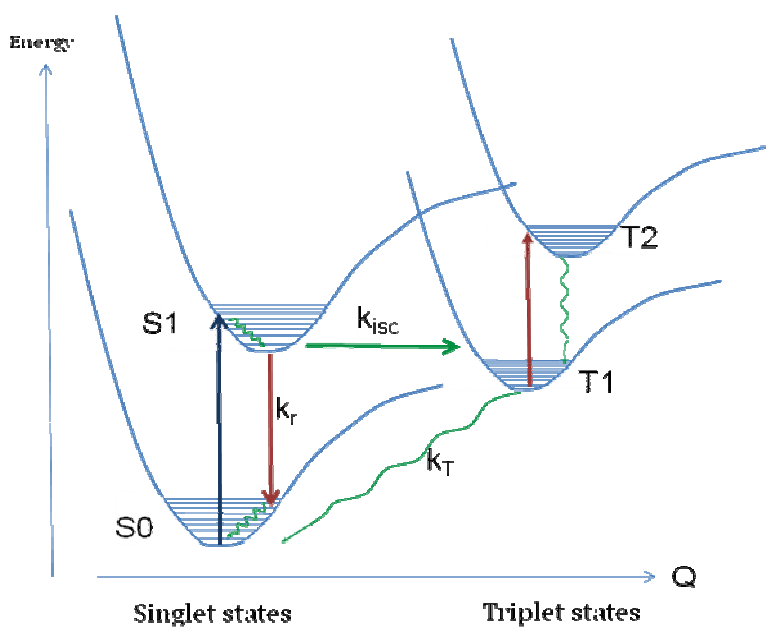

*Figure 4* : Energy level diagram relevant for organic dyes. In organic semiconductors more complex intermolecular (or intrachain) interactions occur (see text). Lasing occurs on the singlet $S_1 \rightarrow S_0$ transition, where SE (stimulated emission) can efficiently shunt the radiative spontaneous emission occurring with a rate $k_r$. Even though it occurs at a slow rate when spin-orbit coupling is weak, intersystem crossing (ISC) causes triplet states $T_1$ to be generated from $S_1$ states. Since their (nonradiative) decay rate $k_T$ is small, they accumulate until they totally hinder SE through a dipole-allowed efficient $T_1 \rightarrow T_2$ absorption (Triplet absorption), or through a nonradiative energy transfer between a molecule in state $S_1$ and another in state $T_1$ (Singlet-Triplet Annihilation).

Absorption cross sections in organics are among the highest of all laser media, typically on the order of $\sigma_{abs} \sim 10^{-16}$ cm². In a neat small-molecular thin film of molecular density $n \approx \rho \times N_A/M = 1.3 \times 6.10^{23}/460 \sim 10^{21}$ cm$^{-3}$ (where $\rho$ is the film density, $N_A$ the Avogadro number and M the molar mass, numerical values here are those of $Alq_3$[71]), incident light will be absorbed over a typical length scale of $L_{abs} = (\sigma_{abs} n)^{-1} \sim 100$ nm. A first consequence is the possibility to make very thin devices, with a strong impact on resonator design (see the corresponding section). A subsequent point is the correlated high gain: as stimulated emission cross section and absorption cross sections are linked and of comparable magnitude, very high optical gains are possible in organic media (gain cross sections around $10^{-15}$ cm² are reported in conjugated polymers[72]).

Another remarkable feature of organic molecules is their very broad fluorescence spectra, leading to a possible tuning of the emission wavelength (thanks for instance to a dispersive element added inside the laser

cavity[73]), and making them capable of ultrashort pulse generation[74]. Furthermore, it is relatively easy to modify the emission spectrum by changing the chemical structure, giving birth to a wide panel of materials emitting from the near-UV to the near-IR, in contrast to inorganic semiconductors where the emitted wavelength is limited by the available materials and lattice-matching restrictions. There are several strategies to chemically tune the emission of a given molecular unit: for instance one may vary the effective conjugation length, *i.e.* the average size of the π-orbital cloud onto the molecule or the segment, which basically defines the color in the same way as the size of a quantum dot relates to its spectrum[75]. An archetypal example is given by the polyacene series, where the adjunction of phenyl rings from benzene to pentacene leads to a redshift of absorption spectrum[76]. Another interesting way of chemical tuning consists in tuning the strength of intramolecular charge transfer (ICT), as illustrated on the unsymmetrical push-pull triarylamine-based compounds shown in figure 5. As the strength of the electron-withdrawing group increases, a higher dipole moment arises in the excited state: when the material is under the form of a neat film or in a polar environment, the Stokes shift increases accordingly and the emission is redshifted.

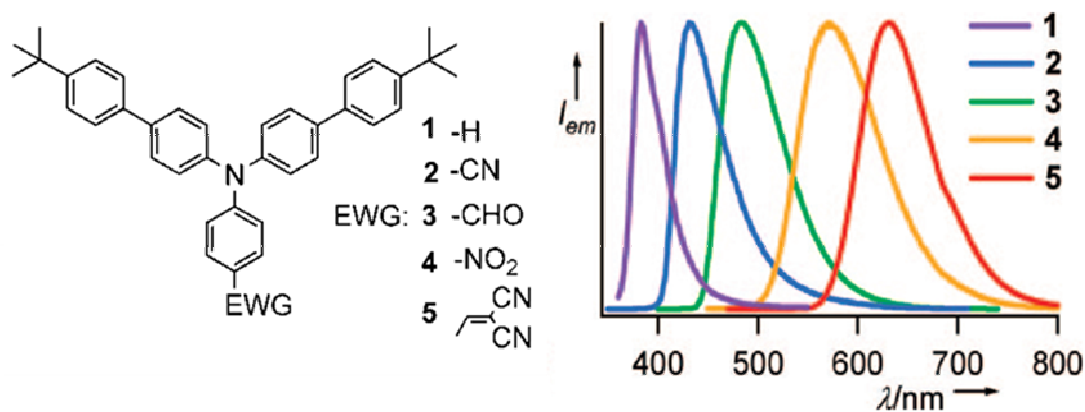

*Figure 5*: Illustration of chemical tuning by substitution of different Electron-Withdrawing Groups (EWG) on a triphenylamino moiety. Right : fluorescence spectra of 100-nm-thick evaporated neat films. Courtesy of E. Ishow. Reprinted with permission from [45]. Copyright 2011 American Chemical Society.

At last, let's mention that photophysics of organic gain media will be different in neat films in contrast to dilute solutions, due to the importance of intermolecular interactions (see section 2.3.)

It is possible to make a short list of the qualities that are required to make a good organic laser material: they pertain both to the fluorophore alone and to its environment (the host matrix for instance):

- Compounds are stable against oxygen and moisture, and photostable against pump photons (pumping at low photon energies greatly relaxes this constraint);
- The quantum yield of fluorescence, measured in the solid state, is high (low quenching, weak π-π stacking, etc.);
- Losses at the laser wavelength are minimized (low reabsorption of the fluorophore and low absorption/scattering losses of the environment);
- The Stokes shift is low (reduces the fraction of pump energy converted into heat, and thus enables to increase the conversion efficiency and photostability. Note that this requirement is in contradiction with the previous point);
- Stimulated emission cross section $\sigma_{em}$ is high (governs the local gain $g = \sigma_{em} \Delta N$ , and then the pump threshold);
- a combination of low intersystem crossing rate, low triplet-triplet absorption cross section, low spectral overlap of optical gain with triplet absorption, and eventually low triplet state lifetime is desirable (all these parameters are often not known but strongly limit the repetition rate and pulse duration obtainable from optically-pumped organic lasers)

- In host-guest systems based on Förster resonant energy transfer, a large Förster radius of several nm is desirable (good overlap of donor emission with acceptor absorption, high cross sections…)

In order to certify a given material for lasing applications, Amplified Spontaneous Emission (ASE) experiments are often carried out to measure the optical gain, especially for thin-film devices when light can be waveguided inside the active material. The pump beam is tailored under the form of a thin stripe: when net gain is present in the film, light emerging from the edge of the stripe exhibits a spectrum much narrower than the fluorescence spectrum, often referred to as "mirrorless lasing". Besides, the edge intensity varies superlinearly with the stripe length, and simple data processing allows extracting the net gain: this technique is referred to as the Variable Stripe Length (VSL) method[46, 77, 78]. Typical gains measured in organic gain media with the VSL technique are in the range $10^0$-$10^2$ cm$^{-1}$. In conjugated polymers, transient femtosecond pump/probe experiments revealed gains that can locally rise up to 2000 cm$^{-1}$ in MeLPPP[79] or 12000 cm$^{-1}$ in Polyfluorene[80].

## 3. A CLASSIFICATION OF MATERIALS: DYES VERSUS ORGANIC SEMICONDUCTORS?

Is there a way to classify these different materials (dyes in inert matrices, host/guest systems, conjugated polymers, small molecules, dendrimers, and so forth) in classes that would be relevant for the laser physicist?

A first distinction can be drawn on the basis of processing techniques, leading to the differentiation between bulk gain media and thin films. *Bulk gain media* are millimetric to centimetric rods that are intended to be set in macroscopic laser cavities after being polished to optical quality[29]; the rods are fabricated by inserting an organic emitter into a monomer (such as MMA) before polymerization[29, 34] or into a glass by a sol-gel process[81]. In contrast, *thin films* are somewhat easier to implement, either through thermal evaporation or solution processing. Small molecules (including dyes) can be evaporated provided that they are neutral, while polymers are too heavy to do so. Solution processing includes techniques such as spin-coating[82], dip-coating, doctor-blading[83], or ink-jet printing[84]. Many conjugated polymers are solution-processable, as well as small-molecular compounds, provided that they are soluble. Some unsoluble polymers may be made solution processable via *e.g.* side chain functionalization or main chain copolymerization with soluble segments. Non-conjugated polymers such as PMMA are also readily solution-processable, making dye-doped polymeric lasers also suitable for thin-film devices.

Beyond this "technological" classification, a long-admitted distinction is traditionally made between "dyes" and "organic semiconductors", and it is useful to wonder what the foundations of such a differentiation are. Following Samuel *et al.* [22], three properties are often argued to erect a division: lasing OSCs, as opposed to dyes, (i) have a high photoluminescence quantum yield even in neat films, (ii) can be processed under the form of thin film structures and (iii) are capable of charge transport. These criteria are however sometimes ambiguous, as some molecules hardly fall into one category exclusively: they may have high quantum yields of fluorescence in solution (complying with the classical definition of a dye) but keep their luminescent *and lasing* properties in neat films, whereby they acquire semiconducting properties. For instance, the unsymmetrical starbust triarylamine derivatives introduced by E. Ishow *et al.* [45] (shown in fig. 5) exhibit quantum yields of fluorescence that are in the range 0.2-0.8 both in cyclohexane solution and in thin solid films, they can be used as neat films in low-threshold optically-pumped lasers [85], and can also be incorporated in OLEDs under the form of 20-nm thick pure layers, proving that they are capable of charge transport[47]; they can be evaporated but are also solution-processable, for example by spin coating. Such compounds comply with both definitions of dyes and organic semiconductors and highlight the porosity of the barrier between these two classes. Historically, the dyes versus OSC separation was motivated by the admitted goal of achieving electrical pumping : nowadays, many recent advances (described in the following) pertain to optically-pumped devices, in which semiconducting properties are unemployed; in those cases making a distinction based on electrical properties seems irrelevant.

However, a distinction can still be made at the macroscopic scale of the gain medium rather than at the molecular scale, based on the occurrence of intermolecular interactions, and their influence on laser performance. In a first category, including neat films and host/guest systems (which can be identified to OSCs), exciton-exciton annihilation, exciton diffusion, photogeneration of charges and related phenomena [86-89] can quench the excited state population, and make the photophysical description of the laser medium quite different from the simple 4-level picture exposed above. This has been shown extensively in conjugated polymers[87],[90], and this is also true in guest/host systems: for instance in the $Alq_3$:DCM blend, the fast laser dynamics are affected by the Förster transfer dynamics[91] as well as by quenching of the guest (DCM) singlets on photogenerated $Alq_3$ triplets[70]. In contrast, when an emitter is dissolved in a passive matrix, intermolecular effects are lower and laser physics somewhat simpler. In that sense, the difference between dyes in optically-inert matrices and organic semiconductors can be related to the importance of intermolecular interactions on lasing properties.

## 3 LASERS ARCHITECTURES

The positive feedback loop induced by the resonator is one of the basic elements of any laser system. The presence of the resonator imposes two properties to the oscillating laser field. First, it defines the allowed resonant frequencies of the laser within the gain spectrum. These longitudinal modes correspond to the cavity resonances imposed by the phase conservation after a roundtrip in the resonator. Secondly, the spatial characteristics of the output beam are also a consequence of the filtering role of the cavity: only the modes that are self replicating after a round trip could be amplified in the structure and ultimately define the transverse pattern of the laser beam. In every laser, the oscillating condition is that the gain has to overcome the various losses along a roundtrip in the cavity. The choice of the resonator parameters, and especially the control of the losses in the oscillating structure, is then of foremost importance to achieve lasing. The optical cavity also acts as a beam extractor: in a simple two-mirror cavity, one of them (the output coupler) is only partially reflective in order to extract a given part of the laser energy from the cavity. The output coupler reflection coefficient controls in this way the output power of the laser beam.

Laser resonators may have various shapes and architectures, often more complex than the archetypal linear Fabry-Perot cavity or ring cavities, which are more suited to bulk laser gain media. Seminal works on organic lasers were however done with liquid dyes in such macroscopic cavities [92, 93]. This kind of resonators is composed of two mirrors placed around the polymer block or the cuvette containing the organic solution. One of the mirrors is often a diffraction grating: upon simple rotation, it is possible to select which wavelength will be retroreflected into the cavity to tune the laser. These macroscopic cavities usually allow very high output energies – up to the mJ level – usually associated with high thresholds[34].

As highlighted in the previous section, a major advantage of organic semiconductors and dyes embedded into polymeric matrices is their ability to be easily processed under the form of thin films of micrometric thicknesses, through cheap and controllable fabrication techniques. Small-molecular neutral organic semiconductors may additionally be thermally evaporated, a technique which provides an excellent control over thickness and optical quality. The most natural resonator geometry for thin-film organic lasers is consequently the two-dimensional planar waveguide. We will focus in the following paragraphs on those "thin film" lasers, and give a rapid review of some of the most frequently used resonator geometries.

## 3.1 WAVEGUIDES

In a planar waveguide, the resonator axis is parallel to the film plane: the photons travel a long way (several millimetres) through the active medium during a roundtrip, leading to high overall optical gain. In this geometry, the light is waveguided in the high refractive index organic layer, sandwiched between, typically, a low-index substrate (glass or silica) and air. As the typical order of magnitude for an organic layer thickness is around 1 µm, single-mode waveguides are easily achieved. Optical feedback can be obtained in several ways reviewed below.

### 3.1.1.1 FABRY-PEROT WAVEGUIDES

A convenient and low-cost arrangement used in inorganic semiconductor diode lasers to form the cavity mirrors is to cleave the semiconductor to obtain flat facets. Indeed, the crystalline structure of the semiconductor allows clean and almost perfect facets, with a reflectivity as high as 30 % due to the very high refractive index of the active medium. Things are not so simple with organic materials, as the refractive index is typically two times lower, leading to only a few percents of reflectivity. Moreover, it is technically very difficult to obtain good quality edges with organic films because of the amorphous structure of the material.

However, Kozlov *et al.* [56] demonstrated such a Fabry-Perot resonator with a 1-mm long $Alq_3$:DCM layer as the active medium. As expected from the long interaction length between the laser wave and the gain medium, the lasing threshold was low (around 1µJ/cm²) and the quantum slope efficiency remains one of the highest ever reported (70%). However, the spatial quality of the emerging beam is low and its divergence is high because of the subwavelength aperture at the end of the waveguide.

### 3.1.1.2 DISTRIBUTED FEEDBACK LASERS (DFB)

A very attractive way to avoid the "facet drawback" together with increasing the mirrors reflectivity is to use diffractive structures. In this case, a periodic diffractive grating is used to provide feedback with a very high efficiency for a given wavelength range. These gratings can be easily integrated onto planar waveguides and thus avoid the need for cleaved facets. In such structures, the high reflection coefficient ensured by the periodic grating is combined with the potential long interaction between the laser wave and the gain medium to ensure low threshold lasing. One can classify the periodic surface corrugation in two categories, namely distributed Bragg resonators (DBR) and Distributed FeedBack (DFB) structures.

A DFB laser consists of a thin active layer deposited onto a corrugated substrate. The light propagates in the waveguide and is scattered by the periodic corrugation. If the scattered waves combine coherently, a Bragg wave is created in a new direction (figure 6). The full description of such a mechanism requires the use of coupled mode theory and has been described by Kogelnik *et al.*[94]. For a given period of modulation, only a given set of wavelengths will correspond to constructive interferences in the direction opposite to the incident wave and leads to the feedback required for laser operation. The bandwidth (or width of the "stop band" around the central Bragg wavelength) is related to the modulation depth. The central Bragg wavelength is given by the Bragg condition:

$$m\lambda_{Bragg} = 2\, n_{eff}\, \Lambda \qquad (1)$$

where $\lambda_{Bragg}$ is the wavelength of the laser light, $\Lambda$ is the period of the corrugation, and m is an integer representing the order of the diffraction process. $n_{eff}$ is the effective refractive index of the waveguide. For first order diffraction process (m=1), we obtain $\Lambda = \lambda/(2\, n_{eff})$. An issue encountered with such a structure is light extraction through highly reflective Bragg mirrors, together with the high divergence due to the small transverse size of the waveguide. To circumvent both problems, the second order grating (m=2) is of particular interest. Indeed, in this configuration the resonant wavelength is λ = $n_{eff}$ Λ: the grating diffraction condition for this order imposes that some light is diffracted perpendicular to the waveguide plane, leading to surface emission [60, 62] (figure 7a). Such second-order structures can provide an efficient way to extract the light from the cavity, even if second order gratings typically lead to a higher lasing threshold compared to first order gratings [95, 96].

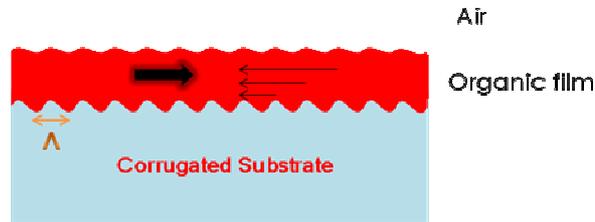

*Figure 6 Possible structure of a DFB laser where the organic film is deposited on a pre-corrugated substrate with a modulation period equal to Λ= mλ/2$n_{eff}$. The light propagating from the left is scattered by the corrugation, and the diffracted waves interfere constructively in the opposite direction, creating a counterpropagating wave.*

Simple one-dimensional DFB organic lasers have been reported in a huge set of materials (both polymers[27, 60, 62, 97] and small molecules [26, 50, 98-101]) covering almost all the visible spectrum. Another advantage of DFB lasers is that their wavelength can be relatively easily tuned by changing the thickness of the organic layer (hence the effective refractive index) or the period of the modulation (see section 5). The concept of DFB laser can also be extended to two-dimensional systems. This type of periodic structures (where either the thickness or the refractive index is modulated in 2D) is referred to as 2D photonic crystal structures[102, 103]. There is a growing interest in such structures, either with square, hexagonal or circular concentric geometries. In such resonators, optical feedback is applied in several directions in the plane of the film. This opens the way toward photonic bandgap engineering where the spectral and spatial characteristics of the laser beam could be controlled with a high degree of accuracy. For example, it has been demonstrated by Samuel's group at St Andrews that a clear enhancement of the spatial beam quality can be obtained through the use of 2D DFB structures [102-105] (even if the far-field pattern generally follows the grating geometry, for example forming a cross in a square-lattice photonic crystal[106]) together with an important decrease of the threshold[105].

### 3.1.1.3 DISTRIBUTED BRAGG REFLECTOR LASERS (DBR)

One potential difficulty with a DFB structure is that the thickness modulation can induce incoherent scattering associated to the surface irregularities, and consequently a high level of losses. Lower threshold can thus be obtained through the use of so-called (DBR) geometries[107]. The term usually applies to a stacking of quarter wavelength layers with alternating high and low indices (in dielectric mirrors for examples). The same principle could be applied in the waveguide plane by corrugating the layer with a periodic pattern with a half wavelength period. The laser then consists of an organic waveguide in which two outer areas are nanopatterned to ensure Bragg reflection [46, 108] (see Figure 7b). Each of the corrugated areas acts as a Bragg mirror, and reflects light for the chosen wavelength. The advantage of such a structure is that the gain medium located in-between remains uncorrugated, which limits losses induced by scattering.

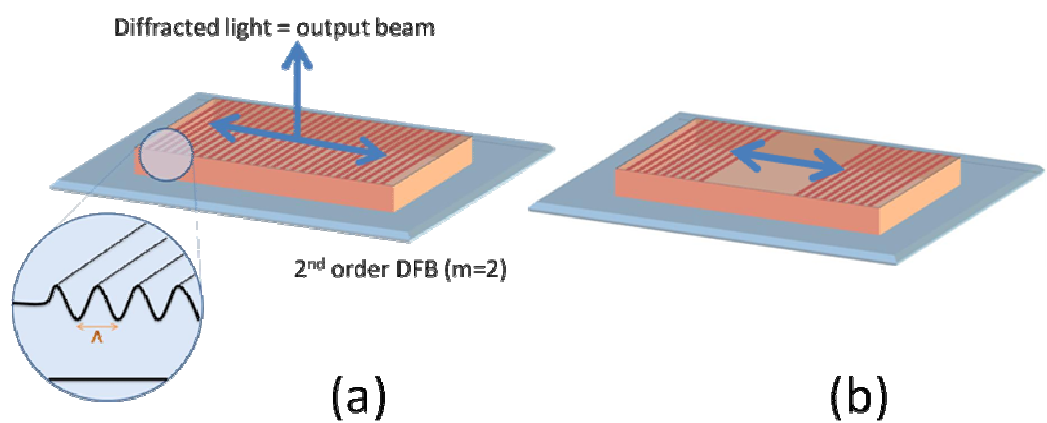

*Figure 7 (a) DFB structure directly engraved onto a polymeric film. If Λ= λ/neff (2$^{nd}$ order grating, m=2) then a diffracted beam (first diffraction order) is emitted in the direction perpendicular to the film plane (b) DBR structure with no corrugation between two Bragg mirrors.*

## 3.2 MICROCAVITIES

### 3.2.1.1 ORGANIC VERTICAL CAVITY SURFACE EMITTING LASERS

Another natural arrangement for an organic laser, inspired by stacked organic light emitting diodes, consists in a planar sandwich of the organic material between two mirrors to form a Fabry-Perot microcavity in the vertical direction (resonator axis perpendicular to the film plane, see figure 8a). This kind of configuration has already been successfully used for crystalline solid-state lasers and for inorganic diode lasers where it is known as Vertical Cavity Surface Emitting Lasers (VCSELs). The OVCSEL (Organic VCSEL) geometry was the one used by Tessler *et al*. [6] to demonstrate optically-pumped organic lasing in PPV. Using a highly reflective broadband DBR mirror and a partially transmitting silver mirror as an output coupler, very high quality factor Q could be achieved[109]. The 100-nm-thick PPV layer was simply spin coated onto the DBR mirror and the silver layer was evaporated on top. Only a few longitudinal modes are supported by a microcavity, and when the pump power exceeds the lasing threshold (200 µJ/cm² in this case), a single mode dominates the emission spectrum of the laser. Following this pioneering work, several planar microcavities with various organic media were reported. The microcavity Q-factor (and thus the lasing threshold) can be improved by sandwiching the polymer between two DBR mirror[110], or by directly growing the top DBR mirror at low temperature onto the polymer[111]. Another option is to change the lasing medium to use small molecules instead of polymers. Lasing with the archetypal $Alq_3$: DCM blend in a DBR/metal [109] or DBR/DBR [91] microcavities were reported. Recently, Sakata *et al.*[112] realized such a microcavity laser with a cheap and compact UV laser diode as a pump source.

### 3.2.1.2 MICRORINGS, MICRODISKS AND MICROSPHERES

The easy processing of organic materials allows for the realization of new geometries that are impossible to obtain with classical inorganic semiconductors. A first example is the microring laser, where a polymeric film is deposited around an optical fiber or even a metallic wire[113] (figure 8b). This can simply be achieved by dipping the fibre into a solution of the polymer[114]. The waveguide created around the fibre forms a ring microcavity where the light bounces through total internal reflection on the interface between the polymer and the surrounding medium. As the diameter of the fibre is typically around several micrometers, the light travels through a long path to make a roundtrip, which is associated to a complex mode pattern, mixing whispering gallery modes and waveguide modes[115]. As a consequence, very low thresholds can be achieved with such structures [116] (around 1µJ/cm²). The pumping of these lasers can be done transversally or through the core of the fibre, leading to a more uniform pumping pattern[117].

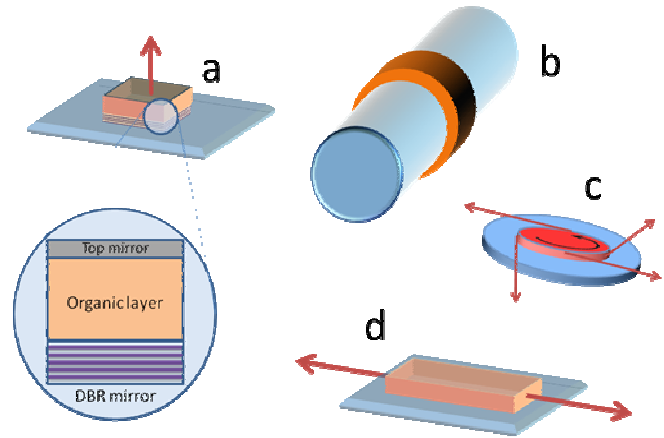

*Figure 8* : Different resonators used for organic semiconductor lasers : (a) vertical Fabry Perot microcavity, (b) microring resonator, (c) microdisk resonator, (d) Planar Fabry Perot Waveguide.

In a similar way, microdisks lasers [19, 116] can be realized by lithographically patterning and etching an organic film to form circular disks of several micrometers of diameters (figure 8c). The microdisk supports whispering gallery modes comparable to those observed in microring lasers, but also possible Fabry-Perot resonances. With more complex shapes[118] (squares, pentagon, stadiums…) a wide variety of mode patterns are addressable, which offers a tool to study quantum chaos or chaotic lasing. The regular comb-like spectrum obtained even with complex shapes enables inferring the dominant periodic orbits in such structures[119] (see figure 9.)

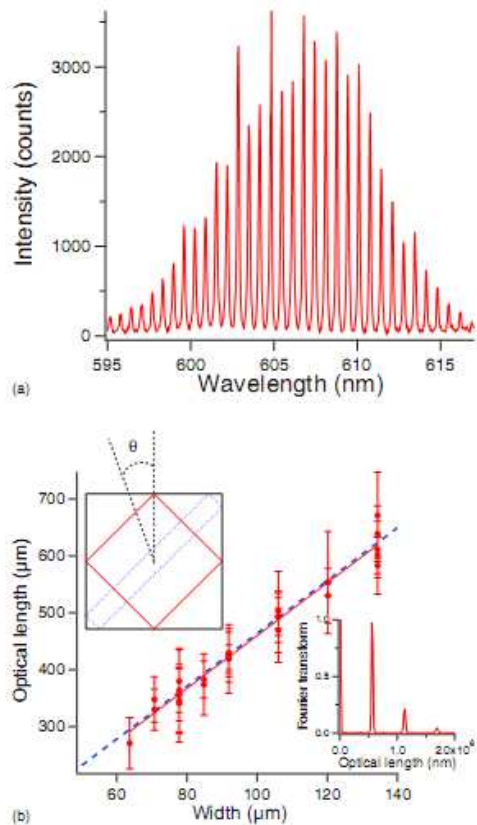

*Figure 9* : optical spectrum obtained from a square-shaped PMMA:DCM microlaser (square width W=150 µm), from ref.[119] (b) : a Fourier transform (inset) of the spectrum allows inferring directly the length of the periodic orbit, which as expected is found to vary linearly with the square dimension W. Reprinted figure with permission from [119]. Copyright 2011 by the American Physical Society.

Finally, the microsphere structure[120] is the 3D generalization of the microdisk. It can be formed by superimposition and melting of several microdisks on a lyophobic surface to form droplets. A common property shared by these geometries is the messy output, since light is emitted in many radial directions. This unusual output may be tailored upon playing on the shape of the resonator[118].

### 3.3 THE VECSOL CONCEPT

A major drawback shared by DBR/DFB, microdisks/rings and microcavity organic lasers is the asymetric and/or highly diverging profile of the output beam, which strongly limits the potential applications of such laser sources. *A contrario*, circular and diffraction-limited beams (*i.e.* with the minimum beam divergence obtainable from a beam of given transverse size, governed by diffraction laws) can be obtained obtained with external bulk resonators using dye-doped polymer blocks but at the expense of compactness and with longer and more complex fabrication processes than organic thin films. A recently demonstrated strategy to obtain a diffraction-limited circular output beam from an organic thin-film laser is the Vertical External Cavity Surface-emitting Organic Laser (VECSOL)[121] which is the organic counterpart of the inorganic VECSEL[122]. It is composed of a plane mirror coated with a thin film of the organic material and a remote concave mirror to close the cavity. The macroscopic (up to several centimeters) cavity defines the spatial geometry of the laser mode and gives birth to a $TEM_{00}$ diffraction-limited transverse profile (see figure 10). This laser architecture also leads to record optical conversion efficiency for thin-films organic lasers (more than 50%[123]) together with respectable output energies (> 30 μJ). This contrasts with the relatively low slope efficiencies obtained with DFB structures (usually a few percents or less).

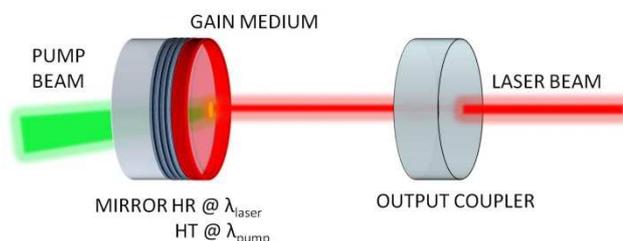

**Figure 10** : *a Vertical External Cavity Surface-emitting Organic Laser (VECSOL)[121] . The left mirror is transparent for the pump wavelength and highly reflective for the laser radiation. The output coupler radius-of-curvature and reflectivity are optimized to yield the highest output energy in a single transverse mode.*

## 4  PROSPECTS FOR ELECTRICAL PUMPING

Soon after an optically-pumped organic solid-state semiconductor laser was demonstrated in 1996[7], an electrically-pumped organic laser, or "organic laser diode", became an important breakthrough worth to be targeted, potentially opening the way to a new class of extremely compact, tunable, cheap, flexible lasers: it remains today a hot topic as no such organic laser diode has been published yet. In 2000, Schön *et al.* [124] reported such an achievement in a tetracene crystal, but this paper (and others by the same author) was later retracted[125], a story which unfortunately entered the history of scientific misconduct rather than the history of laser breakthroughs. Other claims of electrically-driven stimulated emission have been reported during the XXI[th] century: Yokoyama *et al.*[126] observed a spectral narrowing of the edge emission of an OLED as well as an intriguing superlinear dependence of the emission with the electrode length, which was attributed to the presence of stimulated emission. Tian *et al.*[24] observed the same features but suggested that a misalignment of the collecting optics could be the reason for the observed superlinear dependence instead of gain. In both cases, the spectral narrowing can be well accounted by a resonant leaky mode at the cut-off frequency[126]. Another intriguing claim of lasing under electrical pumping in a microcavity OLED was reported by Liu *et al.* [23] with an unrealistically low threshold. Those recent reports, however, are probably not lasing, as shown and discussed by Samuel *et al.*[25] It can be added that stimulated emission is claimed in ref[126] under CW excitation, which is a strong indication against lasing : CW operation under optical pumping would be actually quite a breakthrough in itself (such a report by Nakanotani *et al.*[127] is questionable for the same reasons and is probably only due to a resonant leaky mode).

The challenges to realize direct electrical pumping of a laser device are numerous and complex [70, 128, 129], and are much better understood now than 15 years ago. Quite unfortunately, the astonishing progresses made in OLED technology in the past ten years, in terms of luminance and efficiency, are not straightforwardly transferrable to organic lasers. A good illustration of a breakthrough for OLEDs that is not useful for lasers is triplet energy harvesting through the use of phosphorescent materials[130]. This new class of materials led to a ~ fourfold increase of efficiency as 75% of the excitons created by carrier recombination are triplets. However, phosphorescence is not suitable for lasing (at least attempts to observe ASE in highly-efficient phosphorescent emitters Ir(ppy)$_3$[131] or Btp$_2$Ir(acac)[132] have failed) since the excited state absorption band $T_1 \rightarrow T_n$ (n>1, a dipole allowed transition associated with high absorption cross sections $\sigma_{T-T}$) spectrally overlaps the $T_1 \rightarrow S_0$ emission band, associated with low stimulated emission cross sections.

The key difference between an OLED and a laser is that lasing requires net gain (*i.e.* light amplification surpassing all kinds of losses, both due to the material and the environment), whereas fluorescent emission is a linear threshold-less phenomenon: because π-conjugated systems have short excited state lifetimes (~ a few ns), a high exciton recombination rate will be required to run a laser, much more than in a standard operating OLED. It is instructive to seek an estimation of what would be the current density threshold of an organic laser diode if the losses were the same as in an optical pumping configuration. Starting from optically-pumped DFB laser data, a value around 100 A/cm² can thus be sketched: this estimation was obtained from a DCM:Alq$_3$ active medium[133] sandwiched between two cladding layers of Alq$_3$, as well as from a PPV polymer laser[21]. Importantly, this figure has to be understood as a *lower limit*. Indeed, higher current densities of the order of the kA/cm² have been achieved in pulsed OLEDs[134] or in light-emitting field-effect transistor (LEFET) devices[135] without ever observing lasing, and values as high as 128 kA/cm² have been reached in a thin film of copper phtalocyanine[136]. It makes clear that electrical driving brings extra-losses, due to the presence of metallic electrodes, and also because of absorption and quenching attributed to charge carriers (polarons) and triplet excitons. We briefly review these aspects below.

First, the electrodes used for electrical injection are problematic. The lowest thresholds in optically-pumped devices have been obtained with waveguide lasers (such as DFB), because they enable a long interaction with the gain medium, are consequently a natural design choice for a future organic laser diode. However, the guided mode leaks outside the high-index active part and overlaps with the absorbing metallic electrodes, a feature which is amplified by the fact that all organic films tend to have similar indices of refraction, not helping optical confinement. As a result, the threshold is greatly increased, as it has been shown by optically pumping a structure with passive metallic contacts[137]. It is possible to carefully design the contact geometry so that the overlap between the optical mode and the electrodes is reduced as far as possible[138], for instance by putting the contacts in the node of a TE$_2$ waveguide mode [139], by using thick charge-transport conducting polymers to keep the mode away from the electrode [140], or by using transparent conductive oxides for both electrodes[141, 142].

The second and troublesome source of losses associated with electrical pumping is the presence of the charge carriers themselves (polarons) which have broad absorption bands overlapping the emission and can then absorb laser photons or quench singlet excitons[143]. Little quantitative information was available up to now about polaron absorption: Rabe *et al.* [144] made a precise and unambiguous measurement of a polaron absorption cross section in a hole-transporting spiro bifluorene (S-TAD) compound and found a low value (σ < 2.6.10$^{-18}$ cm² between 560 and 660 nm), confirming the idea that polarons more certainly interact through polaron-exciton quenching [86] than through direct photon absorption. Both the electrode and the polaron issues are in fact directly related to the low mobility of organic semiconductors. Indeed, the mobilities of amorphous organic semiconductors are in the range 10$^{-5}$-10$^{-2}$ cm².V$^{-1}$.s$^{-1}$, which means that achieving a high current flow will mobilize a high number of polarons. A low mobility also means a high resistivity, so that electrodes must be set close to each other to provide a high electric field. A promising laser architecture in this context is the Organic Light-Emitting Field-Effect Transistor (OLEFET)[145], which shows potential for solving the electrode and the polaron problems simultaneously. The key idea is that the field-effect mobility is higher than the bulk mobility measured in a planar OLED structure[146], so that higher current densities can be obtained (values in the range of several kA/cm² have been reported in OLEFETs[135, 147]. This configuration also makes it possible to position the emission zone far away from the metallic contacts. Recently, optically pumped lasing was demonstrated in a LEFET structure with a DFB grating[148]. The structure and the lasing results are detailed in figure 11. Although OLEFETs are now able to outperform the equivalent OLEDs[149], no laser action has been shown yet.

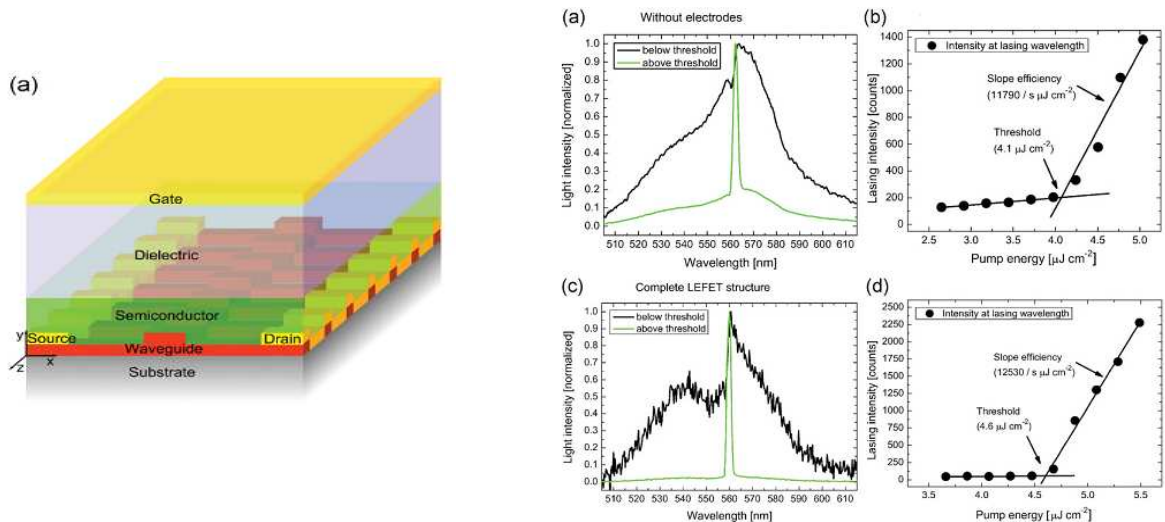

*figure 11* : Left : a Light-emitting field-effect transistor (LEFET) structure incorporating a DFB grating for lasing applications. Right : Spectra and laser curves of the structure without (on top) and with electrodes (bottom). Courtesy of Henning Sirringhaus. Reprinted with permission from [148]. Copyright 2011 John Wiley and Sons.

The presence of triplet excitons is the third and last impediment towards organic laser diodes, and it is certainly the most difficult to overcome. Triplets are more abundant than singlet excitons under electrical excitation: firstly because of the 3:1 unfavourable creation probability ratio, and secondly because their much longer lifetime (~ ms[150]) makes them piling up, at least in a long-pulse or high repetition-rate regime (which makes short pulses always preferrable[151].) Under optical pumping, triplets are only created through intersystem crossing, and are not deleterious provided that the device is operated with short pump pulses. Triplets cause issues because they will absorb laser photons (Triplet Absorption or TA, meaning increased cavity loss), and/or because they will quench singlets through a Förster-type nonradiative dipole-dipole interaction referred to as Singlet-Triplet Annihilation (STA)[70]. These effects are not radically different in nature since they are both related to the triplet-triplet absorption cross section $\sigma_{T-T}$ and the spectral overlap between the singlet emission band and the triplet excited state absorption band $T_1 \rightarrow T_n$ (the same feature that makes organics unsuitable for CW lasing). Contrary to polaron-related phenomena, triplet issues can be studied under optical excitation with long-pulse pumping. Based on the observation of complete lasing suppression a few tens of ns after pump turn-on in $Alq_3$:DCM and BCzVBi:CBP lasers, Giebink *et al*. [70] concluded that STA between guest singlets and host triplets was the dominant mechanism, compared in particular to TA. Lehnhardt *et al.* [63] showed conversely that STA and TA both play an important role in a polymer-polymer F8BT/MEHPPV guest-host system, based on experiments revealing that lasing persists for only a few nanoseconds, regardless of the pump duration. Numerical simulations[70] drove to the conclusion that, even in an ideal case where polaron quenching would be totally absent, triplets hinder lasing in a guest-host system at any current density.

As a conclusion, 15 years after the first demonstration of an organic semiconductor laser, achieving an organic laser diode seems as challenging as it was at the very beginning, but the bottlenecks are now much better understood. The additional losses brought by electrical excitation are now well identified and are being assessed quantitatively. For instance, Wallikewitz *et al.*[140] achieved a direct measurement of all the current-induced losses, by measuring the impact of a current on the threshold of an optically-pumped laser whose active medium is part of a working OLED: they revealed that a current density as low as 7 mA/cm² increased the threshold by 15%. The LEFET idea, while not solving the triplet problem, suggests that real breakthroughs might be expected through a complete reformulation of the traditional concepts inherited from inorganic laser diodes.

## 5. RECENT TRENDS IN OPTICALLY-PUMPED ORGANIC LASER RESEARCH

Facing the numerous challenges exposed above towards the realization of an organic laser diode, some have started to wonder[152] whether the future of compact organic lasers would not be instead an "indirect" electrical pumping scheme, in the sense that charge carriers are not directly injected into the conjugated material but to a laser diode or a LED which in turn would optically pump the material. This application-driven approach reminds us what is the basic ambition for organic lasers: to become cheap, practical, compact alternatives to complex tunable laser technologies. Here, the distinction between a dye (in a polymeric or a semiconductor matrix like $Alq_3$) and an organic semiconductor is useless, since their different photophysical properties (see section 2) are not exploited. Below, we briefly review some recent routes explored towards the improvement of optically-pumped organic lasers. All these developments may not be gathered in a single device (for instance, a low threshold will hardly be compatible with a high output energy) but all participate to the flourishing of the field.

**5.1. The "Indirect electrical pumping" strategy: loosing the constraint on the pump source**

The concept of indirect electrical pumping is to use an electrically pumped light source to (optically) pump an organic laser. This pumping light source must be cheap, compact and efficient, in contrast to expensive flashlamp-pumped Nd:YAG lasers, Nitrogen or Ti-sapphire lasers which have been the most widely used pump sources so far. This approach is then justified by the important progresses made in solid-state coherent light sources technology. Very compact pulsed microchip solid-state lasers have hence been used for the pump[26, 106, 121], leading to organic sources having the size of a small shoebox. A microchip laser remains however a complex light source, composed of an inorganic infrared laser diode and a laser crystal coupled to a saturable absorber emitting infrared light in a pulsed mode, followed by frequency conversion stages to reach the UV-blue wavelengths suitable to pump most of organic lasers. A major advance in terms of reducing costs and improving compacity was triggered by the release in 1996 by the Nichia corp. of blue InGaN laser diodes. Since then, the output powers have been scaled up and became suitable to directly pump organic lasers[27, 108, 112]. Going still further in simplification, Yang *et al*. demonstrated an organic semiconductor laser pumped by an inorganic incoherent light emitting diode (LED), creating a highly compact organic source [153] (see figure 12). This approach opens the way towards highly integrated hybrid organic/inorganic photonics.

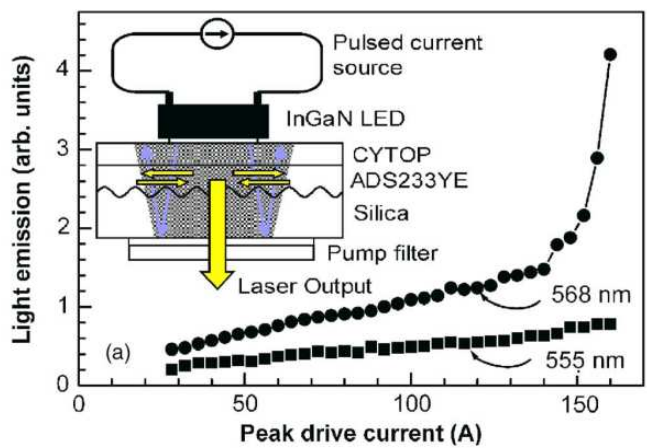

*Figure 12* : Performance of the LED-pumped organic laser proposed by the Ifor Samuel's group. The LED was manufactured by Philips (Luxeon) and driven with short current pulse to reach lasing threshold (details in ref [153]). Inset : scheme of the device : the resonator is a 2$^{nd}$ order DFB grating. Courtesy of Ifor Samuel. Reprinted with permission from [28]. Copyright 2011, American Institute of Physics.

## 5.2. Lowering the threshold

The laser threshold is defined as the pumping energy where the optical gain surpasses the losses over a cavity roundtrip. Achieving a low-threshold laser system is a motivating goal, as it improves the global electrical-to-optical efficiency and lowers the amount of pump energy necessary to drive the laser. With a long term view, an electrically-pumped organic laser would be also obviously demonstrated more easily in a low-loss structure. For a given organic material, the lasing threshold is essentially governed by the interaction length between the laser wave and the gain medium and by the quality of the cavity mirrors. DFB/DBR lasers are intrinsically low-threshold lasers: typical thresholds values for second order DFB lasers are a few µJ/cm² [154] to tens of µJ/cm² [155], but results as low as 200 nJ/cm² [133] or 150 nJ/cm²[50] have been reported for small molecules blends and 40 nJ/cm² for conjugated polymers [9]. First order gratings have the lowest thresholds (see section 3.1.1.2), but at the expense of a degraded output beam quality. A clever approach has been proposed to combine the low threshold induced by first-order gratings with the enhanced beam quality of second-order gratings : with such a mixed structure, threshold as low as 36 nJ/cm² were reported [156].

## 5.3. Extending the wavelength coverage.

π-conjugated systems are essentially destined to emit visible light. At longer wavelengths, nonradiative decay channels compete with fluorescence, resulting in a quantum yield of fluorescence that tends to decrease with the emitted wavelength. On the short-wavelength side, blue or UV emitters suffer from a reduced photostability as small pi-conjugated cores often lack rigidity, and are generally pumped with deep-UV highly energetic photons (although there has been a demonstration of 2-photon pumping of a blue polyfluorene laser [157])

Because of the large potential interest for spectroscopy, substantial efforts have been done to look for efficient UV emitters. Silafluorenes[158] or spiro-compounds[54] are good candidates for this purpose: the lowest lasing wavelength achieved to date directly from an organic semiconductor film is 361.9 nm[159], obtained with a thermally-evaporated spiro-terphenyl film. Telecommunications, biomedical applications (for instance, deep-tissue imaging) and probably in a near future plasmonics (see section 6), will keep on motivating research towards efficient deep-red or infrared gain materials and lasers. Lasing operation has been reported from 890 to 930 nm[160] in a DFB structure, or even at 970nm in a Fabry-Perot configuration, with a commercial dye (LDS 950) doped in a fluorinated polyimide waveguide[161]. The optical gain is low in this case (14 cm$^{-1}$) and the lasing threshold relatively high (600 µJ/cm²). Gain measurements have been reported at a wavelength as high as 1.3 µm in IR1051 dye[162] with a gain of 11 cm$^{-1}$ under 1064-nm pumping.

A radically different solution to the wavelength span limit is to use the visible radiation emitted by organic systems and to convert it into UV or IR using nonlinear optics (that is, using the nonlinear dielectric properties of matter subjected to very intense fields to generate optical harmonics[163].) For a nonlinear process to be efficient, high peak intensities and good beam qualities (i.e. high brightness) must be achieved, which in practice makes this approach easier to implement with external resonators.  S. Chandra *et al*. have obtained tunable UV radiation around 289 nm from a rotating pyrromethene dye plastic disk with external frequency doubling[164]. The same philosophy was applied by Mayer *et al*.[165] to obtain middle-IR radiation around 9 µm from Difference Frequency Mixing of two solid-state dye lasers emitting at 740 nm (oxazine-1) and 803 nm (Rhodamine 800). Recently, this approach was used in a thin-film organic laser with the Vertical External cavity Surface-emitting Organic Laser (VECSOL) concept presented in section 3.3[30]. The authors solved the low intensity problem (associated with thin-film organic lasers) by using an intracavity frequency doubling scheme, made possible by the open structure of the VECSOL. The proposed device is compact (1-cm long) and emits 1 µJ of diffraction-limited tunable UV monochromatic light around 315 nm, from a Rhodamine 640:PMMA active layer spun cast onto one of the cavity mirrors (see figure 13).

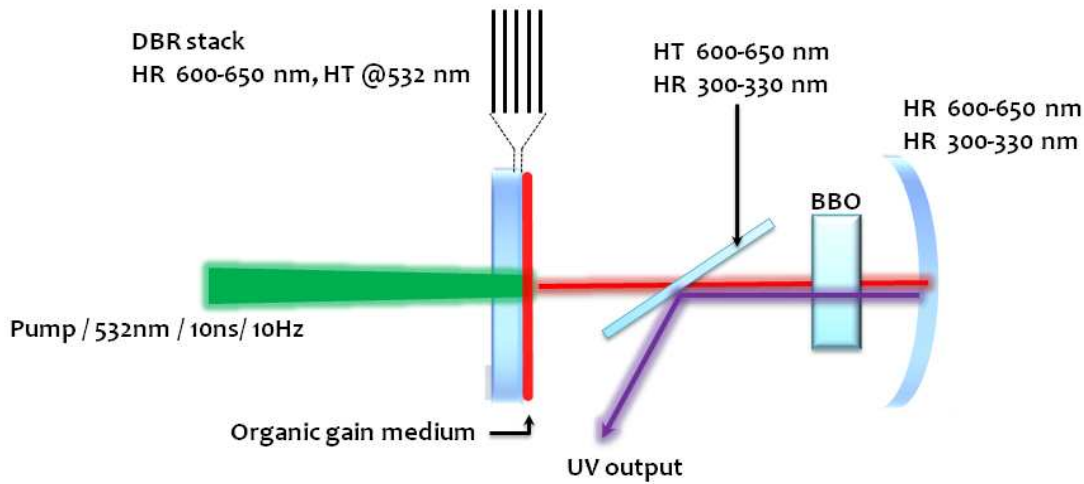

*Figure 13* : Tunable UV organic laser based on a modified VECSOL[30]: a nonlinear crystal is inserted inside the resonator to benefit from the high peak power. A dichroic plate in the cavity allows for the extraction of the UV beam and avoids any accelerated degradation due to the interaction between UV photons and the organic layer. Reprinted with permission from[30] . Copyright 2011, American Institute of Physics.

**5.4. Improving the wavelength agility**

An important aspect that organic lasers must address for practical applications is the wavelength tunability or agility. In a microcavity laser, the emitted wavelength is directly related to the thickness of the active layer. Schütte *et al*. [166] used this property to demonstrate an $Alq_3$:DCM laser continuously tunable between 595 and 650 nm thanks to a wedge-shaped microcavity structure. In this structure the material was evaporated through a rotating shadow mask to a thickness varied between 180 nm and 1850 nm. Similarly, Rabbani *et al.*[121] obtained a 40-nm tunability from a VECSOL, where the spectrum is controlled by the sub-cavity formed by the active layer, upon playing on film thickness variations that arise naturally at the edges of a spun-cast film.

DFB or DBR lasers are easily tunable upon changing the grating pitch, but the tunability is in that case achieved by discrete steps, requiring as many gratings as desired wavelengths[167]. However, one can imagine making the pitch vary in a continuous manner, for instance by using a flexible substrate subjected to a controlled stretching. Using this principle, B. Wenger *et al.*[168] demonstrated a F8BT polymer laser encapsulated in a polydimethylsiloxane (PDMS) matrix with a 20-nm tuning range around 570 nm. P. Görrn *et al*.[169] reported a comparable tuning range in a F8BT/MEHPPV mixture onto PDMS with a superior resistance to stress-induced cracks using a self-organized corrugated PDMS sample, in which ripples are generated thanks to a quite complex procedure allowing for a control of the pitch (see figure 14). In DFB/DBR lasers, the emitted wavelength is also depending on the effective refractive index of the guided mode, as shown in equation (1), offering another route for tuning, either by making an active nanopatterned layer with a wedged shape[170], or by depositing the active layer onto a wedged high index layer[171].

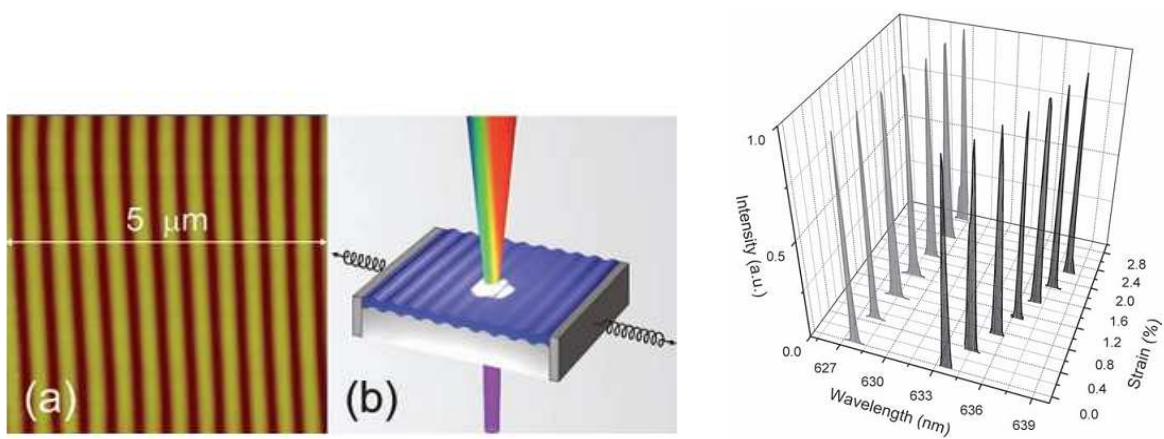

*Figure 14* : an example of wavelength agility on a DFB laser through controlled mechanical stretching, from ref. [169]. Left : AFM image and artist view of the self-organized PDMS grating used as the substrate in these experiments; right : example of wavelength coverage obtained versus applied strain in such laser (gray and black lines compare the tunability obtained with a "cast and cure" grating, obtained by replication onto PDMS od a wafer mater (gray), and self-organized PDMS (black); the active medium is a mixture of F8BT and MEHPPV. Courtesy of P. Görnn. Reprinted with permission from [169]. Copyright 2011 John Wiley and Sons.

**5.5. Improving the conversion efficiency and output power**

Most reports on thin-film solid-state organic lasers emphasize on the spectral narrowing or threshold measurements without giving the obtained output power. The probable reason for that is the difficulty to measure such low output powers or energies with a generally very diverging and non-homogeneous spatial beam profile. In some papers anyway this information is given and interesting slope efficiencies are obtained. For example in an early study Kozlov *et al.* showed a 35 % conversion efficiency (output power/input power) corresponding to a 70% quantum slope efficiency in a Fabry-Perot waveguide configuration (see figure 8.d) with $Alq_3:DCM_2$ used as a gain medium[56]. The maximum output energy in this case was 0.9 nJ. Energies in the nanojoule range are typical for organic semiconductor lasers[105, 106], essentially due to a very small excitation volume. For a large number of applications, the nanojoule energy level could be sufficient anyway, as it correspond to a Watt-level peak power because of the short duration of the emitted pulses. The highest conversion efficiency was obtained recently in a VECSOL structure, where 57% optical-to-optical conversion efficiencies were obtained at the µJ level[123].

**5.6. Improving beam quality**

Beam quality (or spatial coherence) has not been a topic of major interest in early reports of organic lasers. As previously stated (section 3), the spatial beam quality is strongly dependant on the resonator geometry [96, 106] and typical thin-film waveguide laser may have multiple outputs and not-well-defined beams. However, it is possible to obtain a single transverse mode emission: the surface emission with a DFB structure can be near diffraction limited and exhibit relatively low divergence[121], even if a typical fan-shaped beam instead of a desired round homogeneous gaussian beams is observed in these geometries. For edge emission, the small aperture of the waveguide as well as the poor facet quality (especially with spin coated polymers) generally produces highly diverging and inhomogeneous beams. For vertical microcavity structures, it is relatively easy to obtain $TEM_{00}$ emission because of the symmetry of the cavity. However, the output energy is classically very low in those devices. A promising solution to combine relatively high output energies with perfect beam quality is to use an external cavity to control the transverse geometry of the laser beam[121] (see fig.15).

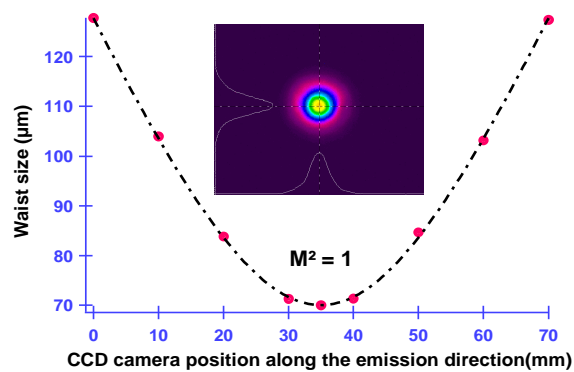

*Figure 15* : *CCD image of the beam profile and M² measurement of a Vertical External Cavity Surface-emitting Organic Laser, from ref [121](M² is defined as the ratio of the divergence to the theoretical diffraction-limited divergence.) Reprinted with permission from[172]. Copyright 2011, Optical Society of America.*

**5.7. Enhancing the lifetime of the devices**

All organic materials tend to degrade rapidly under intense excitation. Lifetime issues have been extensively studied in the field of OLEDs, and thanks to improvements in encapsulating techniques and materials, lifetimes up to 10 million of hours have been demonstrated [57]. Because high peak intensities (and often at short wavelengths) are required to attain threshold in organic lasers, the lifetime issue happens to be more critical than in OLEDs. Even though lifetime measurements are not given in many organic lasers reports, which makes comparisons difficult between materials or structures, a typical maximum is found to be $10^5$ - $10^7$ pulses before the emission intensity decreases by a factor of two (either with microcavities [104] or DFB structures [173] under vacuum). Specific encapsulation schemes can be used to enhance the operational lifetime: it has been demonstrated that a careful encapsulation can enhance the lifetime by a factor of 2500 [174]. The effective lifetime can thus reach hours (depending on the repetition rate), which could be enough for several applications. Another approach is to create nanocomposite materials where the fluorescent units are incorporated into a polymer host acting as an oxygen and moisture barrier [1, 68, 175]. This has also the advantage to create relatively thick self-supported "membranes", which might be disposable and cost-effective.

# 6. IS THERE SOMETHING BEYOND ORGANIC "PHOTON LASERS"? POLARITON LASERS AND SPASERS

In this last section, we highlight two recent results which might open totally new landscapes in the field of organic lasers. They have in common to put into question the very foundation of a laser as a "photon" laser. In these studies, photons are replaced by quasiparticles: either cavity exciton-polaritons or surface plasmons. While the first may be an answer to the organic laser diode problem, the second one puts organic media into the position of playing a role in the generation of optical fields at the nanoscale.

## 1. PHOTONS IN MICROCAVITIES: THE POLARITON LASER

The polariton stands at the crossroad between electromagnetic and matter waves, and its coherent state is thus somewhere between a classical photon laser[176-178] and a Bose-Einstein Condensate[179-181]. In an optical microcavity, this half-light, half-matter quasi-particle is the expression of the strong coupling between excitons and photons inside the cavity[182]. Consequently, the relevant eigenstates of the whole system are not photons or excitons anymore, but a linear combination of both. In the limit of a vanishing photon character and at thermal equilibrium, an exciton Bose-Einstein Condensate is obtained, whereas in the limit of a vanishing exciton character and a nonequilibrium situation the polariton laser is indistinguishable from a classical photon laser [31]. Observation of polaritons at room temperature is tricky because the excitonic part of the polariton dissociates when the thermal energy $k_BT$ is comparable to the exciton binding energy. Interestingly, this binding energy is much higher with Frenkel excitons that best describe organic media than in the Wannier-Mott exciton picture, relevant for inorganic semiconductors (approximately 1 eV versus 10 meV) : polaritons are thus more stable in organic semiconductors. Lidzey *et al.* reported on the first observation of strong exciton-photon coupling in an organic semiconductor cavity more than a decade ago [183] and then suggested that polariton lasing was possible. The first demonstration of polariton lasing in an organic material was actually reported in 2010 in an anthracene single crystal [31]. The 120-nm thick crystal was sandwiched between two DBR and optically pumped by a 150-fs laser. When the threshold was reached (at 320 µJ/cm², presumably lower than for as similar photon laser), a superlinear behavior of the output emission vs. pump power curve was observed together with a reduction of the spectral linewidth, which is said to be a clear laser signature. The spatial profile of the beam also changes to exhibit a structures $TEM_{01}$ shape. According to calculations, direct electrical driving of such structures is theoretically possible, opening a new exciting route to the organic laser diode.

Let's notice that in optical microcavities, another exotic regime has been experimentally demonstrated for the first time in 2010 using dye molecules, namely a Bose-Einstein Condensate (BEC) of photons [184]. In a classical laser, both the state of the gain medium and the state of the light field are far removed from thermal equilibrium; in contrast a BEC of photons consists in accumulating a macroscopic population of photons, which are bosons, in the cavity ground state. The challenge was to achieve thermalization with a process that conserves the number of photons, a prerequisite for BECs, whereas usually photons in a blackbody are absorbed by the cavity walls at low temperatures. Here a BEC was realized at room temperature and thermalization achieved by absorption and re-emission processes in the dye solution. The emission of the device is composed of a nearly-monochromatic peak around 585 nm which is much alike a laser peak, with a $TEM_{00}$ beam above critical point. Although it is clear that this demonstration may essentially serve fundamental physics purposes, the authors say it might be useful to build coherent UV sources.

## 2. NANOPLASMONICS AND "SPASERS"

Plasmonics is a burgeoning field defined as the study of optical phenomena at the nanoscale vicinity of metal surfaces, and offers exciting perspectives of applications in nanophotonics [185], ultrafast nanoelectronics[186], or biomedical optics[187], among others. At first glance, speaking of optics at the nanoscale, and moreover with the involvement of organic gain materials, may seem twice paradoxical: firstly because it is generally admitted that photons cannot be confined to areas much smaller than half their wavelength; secondly because as we saw in section 4, metals cause giant losses, and organic fluorescence near metals is strongly quenched.

Surface plasmons (SP) are collective oscillations of the conduction electrons of a metal. Surface Plasmon Polaritons (SPPs) are of particular interest: they are quasiparticles associated with electromagnetic waves coupled to free electron oscillations that propagate along the interface between a dielectric and a metal with a negative real part of permittivity[188]. These waves have a higher *k*-vector than in vacuum (hence, a lower wavelength) and are evanescent in both the metal and the dielectric, leading to an enhanced electromagnetic field near the interface[189]. However, because of metal absorption, SPPs cannot propagate over long distances, which is a serious impediment to build plasmonic circuitry. Many works of the last decade have then consisted in increasing the propagation lengths of SPPs by appending gain media in the dielectric. There have been several demonstrations of net gain or ASE [190] [191] [192] in plasmonic structures incorporating organic gain materials, which can be dyes in solution[191], in polymer matrices[189] or conjugated polymers[189]. Red (or infrared) dyes are especially relevant here, as a compromise between minimizing metal losses (the longer wavelength the better) and maximizing material gain (in the visible, see section 5.3). M. Gather *et al.*[193] measured a net optical gain of 8 cm$^{-1}$ with the variable stripe-length technique in a Long-Range SPP mode where the gain medium is a mixture of a PPV derivative (MDMO-PPV) with a poly (spirofluorene) polymer. The study of SPP modes also brings information about organics-on-metal quenching: De Leon *et al.* [194] showed that only the nearest molecules (< 10 nm away from the metal) are quenched through a dipole-dipole interaction to the electron/hole pairs of the metals, while most of the molecules located between 10 and 100 nm feed a Short-range SPP mode rather than a radiative photon mode. This idea can be pushed forward to build a "Surface Plasmon Polariton laser" where the oscillating mode is a SPP: such a device was demonstrated by Oulton *et al*. [195] with inorganic CdS nanorod on silver. The device is characterized by a tight sub-wavelength confinement of the optical mode. Recently, Bergman and Stockman[185] proposed a different way of approaching sub-wavelength lasers with the "spaser" ("Surface Plasmon Amplification by Stimulated Emission of Radiation"): it makes use of nanoparticle (or nanolocalized) SPs instead of SPPs, which are the quanta of the collective oscillations of free electrons in a metal nanoparticle whose dimensions are between ~1 nm and the metal skin depth (~25 nm) [32]. When such a nanoparticle is surrounded by a gain medium, a macroscopic population of SPs is expected to build from stimulated emission. The first experimental demonstration of a spaser was provided in 2009 by Noginov *et al.*[32], who used 44-nm-in-diameter spheres composed of 14-nm-in-diameter gold cores surrounded by OG-488 dye-doped silica shells as the gain medium. The nanospheres were optically pumped at 488 nm and emitted a narrow line at 531 nm, which was assigned to the emission of dye molecules into a surface plasmon mode instead of a photon mode. The interest of SPASERs lies in their ability to generate electric fields oscillating at optical frequencies in *nanolocalized* areas[196]; this may be useful for microscopy (such as SNOM or fluorescence imaging with single molecule sensitivity), nanoscale lithography[185], or serve as a basis for ultrafast nanoamplifiers that have the potential to be the future building blocks of optical computers just as MOSFET transistors are the basis of today's electronics.

The nature of the gain medium in nanoplasmonic devices is, at this time of early demonstrations of physical concepts, quite unimportant, and other media such as quantum dots or inorganic semiconductors may find more useful in some applications. However the first spaser demonstrated by Noginov used organic dyes to supply gain; and it is reasonable to think that organics, as *high gain* media in the visible, added to their versatility in terms of processing techniques, will play a significant role within the next years for the development of nanoplasmonics.

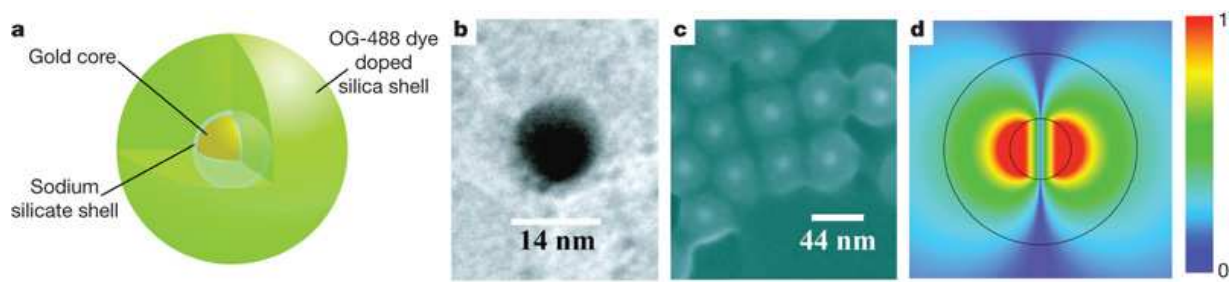

*Figure 16 a*, Diagram of the hybrid nanoparticle architecture (not to scale), indicating dye molecules throughout the silica shell. *b*, Transmission electron microscope image of Au core. *c*, Scanning electron microscope image of Au/silica/dye core–shell nanoparticles. *d*, Spaser mode (in false colour), with $\lambda$= 525 nm and Q = 14.8; the inner and the outer circles represent the 14-nm core and the 44-nm shell, respectively. The field strength colour scheme is shown on the right. Courtesy of Mikhaïl Noginov. Reprinted by permission from Macmillan Publishers Ltd, Nature ref.[32] , copyright 2011.

# Conclusion

In this topical review, we walked around the field of solid-state organic laser research. The available gain materials ("dyes" or "organic semiconductors", depending on the strength of intermolecular interactions in the film and how they influence lasing) are plethora and now cover a wide spectrum from the UV to the IR. Claimed as the main goal to reach as soon as the first optically-pumped organic semiconductor laser was announced in 1996, the electrically-pumped organic laser remains unrealized at the beginning of 2011, and the task seems more complicated than previously thought, because of many additional losses brought by the injected current, that have been identified during the last 15 years and are now being quantified properly. Among the three majors issues — electrode losses, polaron absorption/quenching and triplet absorption/quenching — the last issue is the most problematic and the only one for which no real solution has been proposed to date. However proposals such as lasing Light-emitting Field-effect transistors are promising and the recent demonstration of an organic polariton laser may open new exciting paths. But organic lasers have a great deal to offer under optical pumping, as established for instance by the "indirect pumping" strategy which consists in using cheap LEDs to pump organic lasers. We have reviewed these recent works that aim at making organic lasers more compact, more efficient, more tunable or with a wider span of wavelengths. Aside from these application-driven researches, organic gain media have also found a fertile playground in more fundamental studies involving microcavities, like polariton lasing or Bose-Einstein condensation of photons. At last, they may find a promising future in nanoplasmonics: organics have been used as gain providers for Surface Plasmon Polaritons in plasmonic waveguides, or recently were the basis of the first "spaser", a nanoscale generator of coherent optical fields, with many envisioned applications in bio-imaging or nanoelectronics. These developments continue to trigger fruitful relations between chemists and laser physicists and prove that there are still exciting new challenges in light-emitting molecular materials beyond organic light-emitting diodes.

*Acknowledgments : the authors are very grateful to Elena Ishow and Alain Siove for stimulating and fruitful discussions.*